\documentclass[]{icrc}

\usepackage{times}
\usepackage[]{graphicx} 

\begin{document}

\title{Search for gamma-rays above 10 TeV from Markarian 421 in a high
state with the CANGAROO-II telescope}

\author[1]{K.~Okumura}
\affil[1]{Institute for Cosmic Ray Research, University of Tokyo, Kashiwa, 277-8582 Chiba, Japan}
\author[1]{R.~Enomoto}
\author[2]{A.~Asahara}
\affil[2]{Department of Physics, Kyoto University, Sakyo-ku, Kyoto 606-8502, Japan}
\author[3]{G.V.~Bicknell}
\affil[3]{MSSSO, Australian National University, ACT 2611, Australia}
\author[4]{P.G.~Edwards}
\affil[4]{Institute of Space and Astronautical Science, Sagamihara, Kanagawa 229-8510, Japan}
\author[5]{S.~Gunji}
\affil[5]{Department of Physics, Yamagata University, Yamagata, Yamagata 990-8560, Japan}
\author[6]{S.~Hara}
\affil[6]{Department of Physics, Tokyo Institute of Technology, Meguro-ku, Tokyo 152-8551, Japan}
\author[7]{T.~Hara}
\affil[7]{Faculty of Management Information, Yamanashi Gakuin University, Kofu,Yamanashi 400-8575, Japan}
\author[8]{S.~Hayashi}
\affil[8]{Department of Physics, Konan University, Kobe, Hyogo 658-8501,
Japan}
\author[9]{C.~Itoh}
\affil[9]{Faculty of Science, Ibaraki University, Mito, Ibaraki 310-8512, Japan}
\author[1]{S.~Kabuki}
\author[8]{F.~Kajino}
\author[1]{H.~Katagiri}
\author[2]{J.~Kataoka}
\author[1]{A.~Kawachi}
\author[10]{T.~Kifune}
\affil[10]{Faculty of Engineering, Shinshu University, Nagano, Nagano 380-8553, Japan}
\author[2]{H.~Kubo}
\author[6]{J.~Kushida}
\author[8]{S.~Maeda}
\author[8]{A.~Maeshiro}
\author[11]{Y.~Matsubara}
\affil[11]{STE Laboratory, Nagoya University, Nagoya, Aichi 464-8601, Japan}
\author[12]{Y.~Mizumoto}
\affil[12]{National Astronomical Observatory of Japan, Mitaka, Tokyo 181-8588, Japan}
\author[1]{M.~Mori}
\author[6]{M.~Moriya}
\author[13]{H.~Muraishi}
\affil[13]{Ibaraki Prefectural University, Ami, Ibaraki 300-0394, Japan}
\author[11]{Y.~Muraki}
\author[7]{T.~Naito}
\author[14]{T.~Nakase}
\affil[14]{Department of Physics, Tokai University, Hiratsuka, Kanagawa 259-1292, Japan}
\author[14]{K.~Nishijima}
\author[1]{M.~Ohishi}
\author[15]{J.R.~Patterson}
\affil[15]{Department of Physics and Math. Physics, University of Adelaide, SA 5005, Australia}
\author[6]{K.~Sakurazawa}
\author[1]{R.~Suzuki}
\author[15]{D.L.~Swaby}
\author[6]{K.~Takano}
\author[5]{T.~Takano}
\author[2]{T.~Tanimori}
\author[5]{F.~Tokanai}
\author[1]{K.~Tsuchiya}
\author[1]{H.~Tsunoo}
\author[14]{K.~Uruma}
\author[5]{A.~Watanabe}
\author[9]{S.~Yanagita}
\author[9]{T.~Yoshida}
\author[16]{T.~Yoshikoshi}
\affil[16]{Department of Physics, Osaka City University, Osaka, Osaka 558-8585, Japan}

\correspondence{K.Okumura (okumura@icrr.u-tokyo.ac.jp)}

\runninghead{K.~Okumura et al. : Search for gamma-rays above 10~TeV from
Mkn 421}
\firstpage{1}
\pubyear{2001}

\titleheight{15cm} 

\maketitle

\begin{abstract}
A preliminary result from Markarian 421 observations in the energy region
above 10 TeV with the CANGAROO-II 10 m telescope is presented.
In January 2001, the HEGRA group reported that Markarian 421
had become very active, with flux levels up to 4 times that 
of the Crab Nebula. As a result,
we observed Mkn 421 during six nights from January 24th
to February 1st, and four nights from March 1st to 4th.
Observations were carried out using the very large zenith angle
technique ($\sim$70 degree) and the energy threshold is estimated
from Monte Carlo simulations to be around 10 TeV.
We have detected gamma-ray emission in this energy range.
\end{abstract}

\section{Introduction \label{section:introduction}}

Markarian 421 (Mkn 421, $z$=0.031) was the first extra-galactic 
source from which TeV gamma-ray emission was detected
\citep{punch92}.
Extensive measurements of the gamma-ray flux in the TeV energy region 
have been carried out by several
groups using the ground-based imaging \v{C}erenkov telescope
\citep{krennrich99, aharonian99, piron01}.
The energy spectrum up to 10~TeV has been measured by observations at
large zenith angles, up to 60 degrees,
and can be well described by a power law with a spectral 
index $-$2.5$\sim-$3.0.

It is generally believed that TeV gamma-rays from extragalactic sources
will suffer from the absorption due to
photon-photon collisions with inter-galactic infrared radiation
\citep{gould67,stecker92}.
However, at present, no cutoff due to this absorption 
has been detected in the Mkn 421 spectrum.
In order to determine meaningful constraints on this expected cutoff,
it is important to extend measurements of the spectrum to energies
above 10~TeV.

In this paper, a preliminary result from Mkn 421 observations
with the CANGAROO-II 10 m telescope is presented.
We observed this target at very large zenith angles of about 70 degree
and searched for gamma-rays above 10~TeV.
A Monte Carlo study of this observation mode is also described.

\section{Observation of Mkn~421}

The CANGAROO-II telescope is located near Woomera, South Australia 
(136$^\circ$47$'$E, 31$^\circ$06$'$S).
A detailed explanation of the CANGAROO-II telescope 
is given elsewhere 
\citep{tanimori99,morim99,morim00,morim01,kubo00,kawachi01}.

%
%
The telescope was pointed 
in the direction of Mkn421 for 
six nights from January 24th to February 1st, 
during which two nights were lost due to bad weather,
and four nights from March 1st to 4th in 2001.
Information about these observing periods is 
summarized in Table~\ref{table1}.
Mkn 421 was in a high state during these observation periods.
In the first period, 
the TeV activity was monitored night by night by the HEGRA group
\citep{hegra}.
Their preliminary result shows that flux lies in the range of 
1$\sim$4 times that of the Crab Nebula.
Although there was no HEGRA measurement in the second period,
Mkn~421 was still in a high state according to the X-ray measurements of 
the RXTE/ASM group \citep{rxte}.

%
%
Mkn 421 was observed at zenith angles of
$\sim$70~degrees
as the elevation angle in culmination is 20.7 degrees
at the site of CANGAROO-II telescope.
Observations were scheduled when the source was above 
an elevation angle of 18 degrees.
The observable time amounts to about two and a half hours per night.
Data were taken on moonless and clear nights. 
An OFF source run, displaced in right ascension,
followed or preceded the ON source run.
An event trigger was registered when $3\sim5$ individual pixels
exceeded a threshold of about 3 photoelectrons,
depending on the night sky background.

\begin{table}
\begin{center}
\begin{tabular}{ccc}
\hline
\hline
Date & MJD start & obs. time (hr.) \\
\hline
24-Jan-2001  & 51933.7041 & 1.1 \\
26-Jan-2001  & 51935.7052 & 1.2 \\
27-Jan-2001  & 51936.7050 & 1.6 \\
30-Jan-2001  & 51939.6737 & 2.1 \\
31-Jan-2001  & 51940.6737 & 2.1 \\
1-Feb-2001  & 51941.6777 & 1.9 \\
1-Mar-2001 & 51969.6381 & 0.9 \\
2-Mar-2001 & 51970.6346 & 1.2 \\
3-Mar-2001 & 51971.6144 & 1.6 \\
4-Mar-2001 & 51972.6549 & 0.7 \\
\hline
\hline
\end{tabular}
\end{center}
\label{table1}
\caption{Date, modified Julian day and effective observation hours 
for Markarian 421 observations.}
\end{table}

\begin{figure}[t]
\vspace*{2.0mm} 
\includegraphics[width=8.3cm,bb=28 235 535 598]{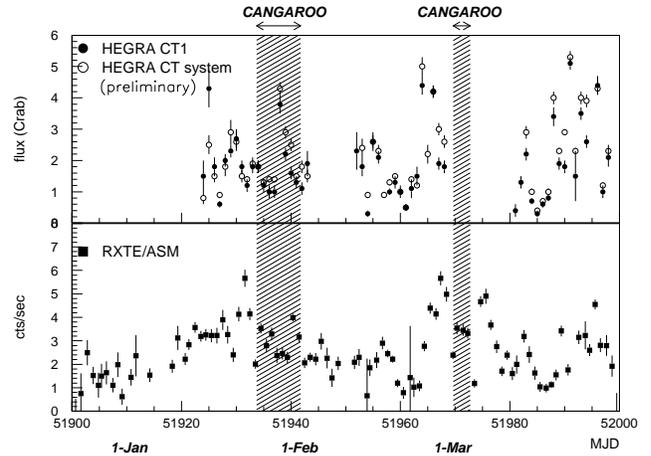} 
\caption{
TeV and X-ray activity measured by the HEGRA and RXTE/ASM groups,
from \citep{hegra} and \citep{rxte}.
The one-day averaged data is shown for lightcurve by RXTE/ASM.
The observation period of the CANGAROO-II telescope is shown 
by the hatched region. 
}
\label{fig:lightcurve}
\end{figure}

\section{Analysis method}

%
%
The ADC and TDC information recorded for individual tubes
are calibrated with the events triggered with a blue LED flasher before
each observation.
Flat fielding of the photomultipliers' gains and 
correction of TDC slewing effect are carried out
using this data.
Camera pixels triggered by sky noise are eliminated 
by the following procedure of image cleaning.
Pixels with pulse heights greater than $\sim$3.3 photoelectron,
triggering within 30~nanoseconds of the timing center of the event
and with more than three neighboring hit pixels are used.
Final acceptance requires more than five hit pixels in each event.

In order to reject data affected by cloud etc.,
data for this analysis were selected based on 
the following criteria:
Data recorded under good sky conditions are used;
The elevation angle of Mkn~421 was above 18.5~degrees; 
and data, which event rate after the image cleaning procedure 
is less than by 2~r.m.s.\ from the averaged rate of the whole data,
are eliminated.
This cut does not exclude any reasonable short-term gamma-ray flare.

As a result, observations totalling
14~hours for ON source and 13~hours for OFF source were selected.
The average elevation angle of the selected data is 20.2 degree
and its r.m.s.\ is 0.6 degree,
thus data which is taken around culmination are selected.

The selection of gamma-ray events is based on the 
standard imaging analysis parameters;
``Width'', ``Length'', ``Distance'' and ``Alpha''  \citep{hillas82}.
In this analysis, 
we adopted the Likelihood method \citep{enomoto01} 
for the selection of gamma-ray events,
instead of the conventional parameterization method.
The discrimination of gamma-ray events is based on 
the probability ratio $R_{prob}$, expressed as follows :

$$
R_{prob} = \frac{Prob(\gamma)}{Prob(\gamma)+Prob(B.G.)}
$$

\noindent
where $Prob(\gamma)$ and $Prob(B.G.)$ are 
calculated probabilities of image shape parameters for
gamma-ray and background events, respectively.
These are the products of individual probabilities for
``Width'', ``Length'' and ``Distance'',
which are derived from the functionized distribution 
using gamma-ray simulations and and OFF source events.
Energy dependence is taken into account in these probability functions.
We adopted a cut of 0.56, as the signal-to-noise (S/N) ratio 
was maximized at this value in the study using simulated gamma-rays and
background events.

\section{Monte Carlo simulation}

\begin{figure}[t]
\begin{center}
\includegraphics[width=8.3cm,bb=50 190 500 625 ]{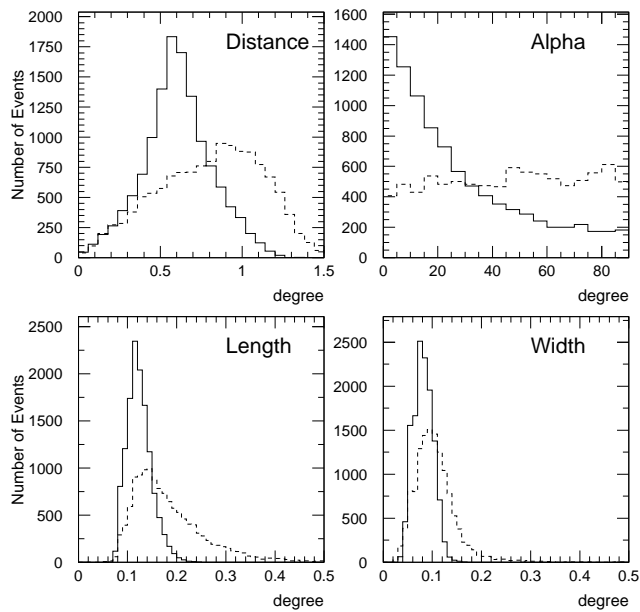}
\caption{Shower parameter distribution of gamma-ray Monte Carlo 
simulation (solid line) and background (dashed line).
The ``Alpha'' distribution is obtained for selected events by
the cut of image parameters.}
\label{fig:parameter}
\end{center}
\end{figure}

\begin{figure}[t]
\includegraphics[width=4.15cm,bb=132 274 416 535 ]{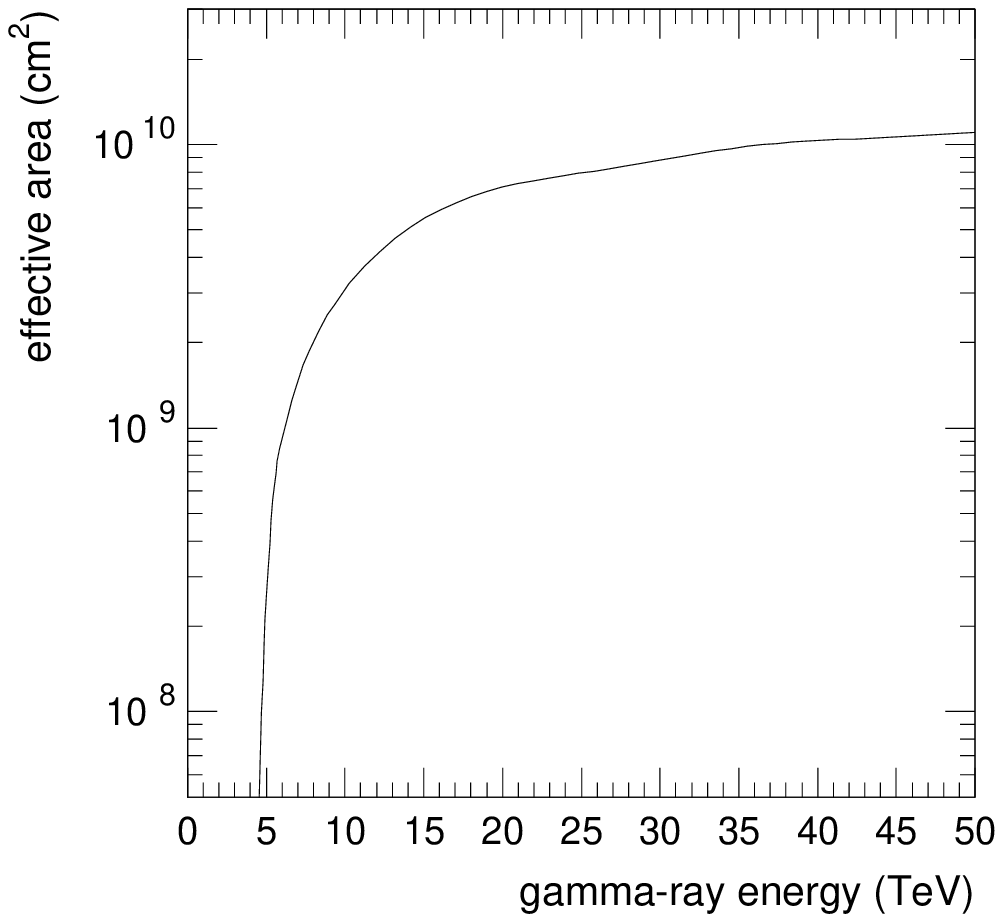}
\includegraphics[width=4.15cm,bb=132 274 416 535 ]{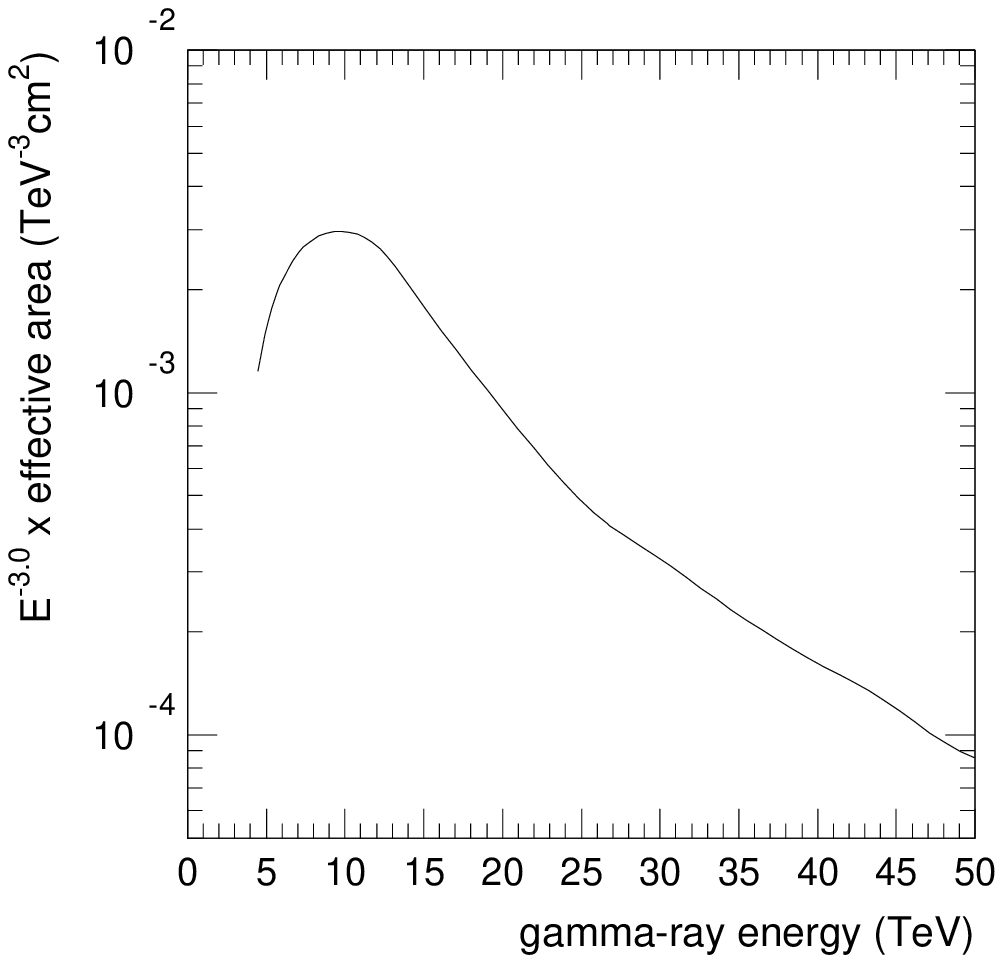}
\caption{Effective area (left) and its product with $E^{-3.0}$ 
spectrum (right). The energy threshold is estimated to be around 9.3 TeV,
which is defined as the maximal point of this distribution. }
\label{fig:effarea}
\end{figure}

%
%
We have carried out a Monte Carlo simulation of
the development of electro-magnetic shower cascade, 
generation and attenuation of \v{C}erenkov photons,
reflection of photons at the mirror surface and 
simulation of electrical hardware.
GEANT simulation code \citep{geant} is used for 
the framework of particle tracking.
Photon attenuation in the atmosphere, mirror reflectivity, 
and quantum efficiency of photomultipliers
are included as a function of \v{C}erenkov photon wavelength.
The measured point spread function and 
mirror reflectivity is used.
Performance of ADC and TDC electronics is also simulated,
including accidental hits due to sky noise.

Gamma-ray events are generated in the energy range
of 3$\sim$50 TeV with the assumed spectrum of $E^{-3.0}$.
The elevation angle of each event is adjusted according to 
the distribution of selected data
so as to reduce the difference in energy scale between 
real data and simulation,
since the energy scale is very sensitive for
large zenith angle observations.

Fig~\ref{fig:parameter} shows the parameter distributions 
obtained from gamma-ray simulations and OFF source events.
For observations at large zenith angle, the shower development point is
relatively farther from the telescope due to the thicker atmosphere.
This implies a higher energy threshold with a broader effective area
\citep{sommers87,tanimori94,tanimori98}.
The shower images focused on camera plane become smaller
and the location of images moves closer to the target direction.
The parameter values become smaller,
compared with those of the observation near zenith, due to these effects.
The resolution of the pointing angle, ``Alpha'', also decreases,
but is still usable.
Naturally the signal-to-noise ratio is smaller than for higher 
elevation observations,
but not so small that it is impossible to discriminate against
cosmic-ray protons.

Fig~\ref{fig:effarea} shows the effective area
and relative event rate for gamma-rays, which is the product of 
the effective area times the assumed spectrum of $E^{-3.0}$.
The energy threshold is defined where the event rate becomes maximal 
and estimated to be 9.3~TeV.
The averaged effective area above this energy threshold 
is calculated to be $5 \times 10^9$ cm$^2$.
It increases to $10^{10}$ cm$^2$ as the gamma-ray energy increases.

\begin{figure*}[t]
\figbox*{}{}{
\includegraphics[width=11.0cm,bb=130 270 416 553 ]{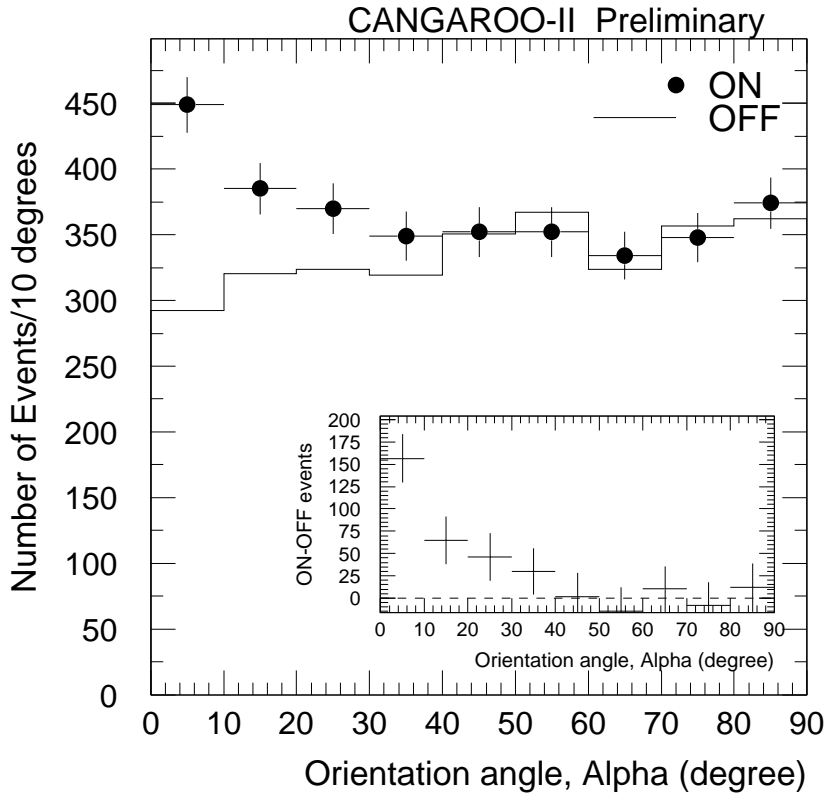}
}
\caption{Image orientation angle ``Alpha'' distribution of Markarian 421
observed by CANGAROO-II 10 meter telescope.
Dots with error bars are ON source data and the solid line is for 
the OFF source data.
The background-subtracted distribution is shown 
in the inserted graph. }
\label{fig:alpha}
\end{figure*}

\section{Result and discussion}

Fig~\ref{fig:alpha} shows the distribution of the image orientation 
angle ``Alpha''. 
The normalization factor between ON and OFF observations is obtained
from the number of events in the ``Alpha'' distributions between
40$^\circ$ and 90$^\circ$.
The gamma-ray signal region is alpha~$<$~20~degrees,
and the excess events after subtraction of the background
is 221 $\pm$ 39 events,
corresponding to a significance of 5.6~$\sigma$.
The broader alpha acceptance is
due to the shrinkage of the shower images as predicted
from gamma-ray simulations at very large zenith angles.

This preliminary result implies the detection
of gamma-rays from Mkn 421 in the energy region of 10$\sim$50~TeV.
A refined analysis will be important for
studies of the infrared radiation absorption,
as discussed in section \ref{section:introduction}.
Furthermore, we have demonstrated that
there is sensitivity for the higher energy region above 10~TeV
from the study of Monte Carlo simulation.
Further analysis may enable constraints to be placed on
the intergalactic infra-red background.

\section{Conclusion}

We have observed Markarian 421 with the CANGAROO-II 10\,m telescope
for 14 hours (ON source data after cuts) while the source was in a
high state at both X-ray and TeV energies
around the end of January and the beginning of March, 2001.
The observations were carried out 
at very large zenith angles, around 70~degrees.
The viability of gamma-ray observations
are verified with Monte Carlo simulations.
The energy threshold and effective area are estimated 
to be 9.3~TeV and  $5\times10^9$ cm$^2$, respectively.
An excess of 221~$\pm$~39 events was found 
using Likelihood method.
This corresponds to a significance of 5.6~$\sigma$ and 
the preliminary detection of gamma-rays in the energy range of 10$\sim$50~TeV.

%

%


\begin{acknowledgements}
We thank the DSC Woomera for their assistance in constructing the
telescopes. 
This project is supported by a Grant-in-Aid for
Scientific Research of Ministry of Education,
Culture, Science, Sports and technology of Japan
and Australian Research Council.
KO is supported for this work by a JSPS postdoctoral fellowship.
\end{acknowledgements}

%
%

%
\end{document}